\documentclass[pre,aps,showpacs,twocolumn,floatfix,superscriptaddress]{revtex4}
\usepackage{graphicx}% Include figure files
\usepackage{bm} % bold math
\usepackage{epsfig}
\usepackage{color}
\usepackage[export]{adjustbox}

\newcommand{\be}{\begin{equation}}
\newcommand{\ee}{\end{equation}}
\newcommand{\bea}{\begin{eqnarray}}
\newcommand{\eea}{\end{eqnarray}}
\newcommand{\ba}{\begin{array}}
\newcommand{\ea}{\end{array}}

\begin{document}

%\preprint{}
\title{Force spectroscopy analysis in polymer translocation}

\author{Alessandro Fiasconaro}
\email{afiascon@unizar.es}
\affiliation{Dpto. de F\'{\i}sica de la Materia Condensada,
Universidad de Zaragoza. 50009 Zaragoza, Spain}

\author{Fernando Falo}
\affiliation{Dpto. de F\'{\i}sica de la Materia Condensada,
Universidad de Zaragoza. 50009 Zaragoza, Spain}
\affiliation{Instituto de Biocomputaci\'on y F\'{\i}sica de Sistemas
Complejos, Universidad de Zaragoza. 50018 Zaragoza, Spain}

\date{\today}

\begin{abstract}
This  paper reports the force spectroscopy analysis of a polymer that translocates from one side of a membrane to the other side through an extended pore, pulled by a cantilever that moves with constant velocity against the damping and the potential barrier generated by the reaction of the membrane walls. The polymer is modeled as a beads-springs chain with both excluded volume and bending contributions, and moves in a stochastic three dimensional environment described by a Langevin dynamics at fixed temperature. The force trajectories recorded at different velocities reveal two exponential regimes: the force increases when the first part of the chain enters the pore, and then decreases when the first monomer reaches the trans region. The spectroscopy analysis of the force values permit the estimation of the limit force to allow the translocation, related to the free energy barrier. The stall force to maintain the polymer fixed has been also calculated independently, and its value confirms the force spectroscopy outcomes.
\end{abstract}


% 05.40.-a  Fluctuation phenomena, random processes, noise, and Brownian motion
% 87.15.A-  Theory, modelling, and computer simulation
% 87.10.-e  General theory and mathematical aspects
% 36.20.-r  Macromolecules and polymer molecules
% 87.15.H   Dynamics of biomolecules
% 87.23.Cc
%\PACS 05.40-a,87.23Cc,89.75-k, 87.17.Aa, 82.20.-w

%Statistical Mechanics, Mean First Passage Time, Noise-induced effects, NES

\maketitle

\section{Introduction}
Translocation of long molecules through membrane nanopores is a common process in living cells. Drug delivery, DNA, RNA, and proteins passage through cell and/or membrane pores, as well as DNA injection and packaging by phage viruses are a few examples of a broad interesting phenomenology~\cite{MetzlerSM2014}.
The passage of polymers through nanopores is also a fundamental problem in those nanotechnological studies that try to emulate the complex biological processes involved in the phenomenon~\cite{li2001,mickler}, and on translocation is based the next generation of DNA sequencing technique~\cite{sigalov2008,Zwolak2008,MenaisSR2016,GolestPRX}.

For these reasons translocation processes is under a deep investigation. With the aim of efficiently describing the complexity of biological matter in a reasonable time, different mesoscopic models for polymer translocation have been introduced~\cite{Meller2003}. 

Different works have treated the translocation as a stochastic diffusion through a single free energy barrier \cite{Sung1996,Muthukumar1999,Muthukumar2011}, which is a function of the excess number of crossing monomers of a bead-sticks chain. In simulation, different models of a single barrier potential, eventually depending on time, have been introduced to depict the overall translocation process of a bead-spring chain~\cite{Pizz2009,Pizz2010,Pizz2013}. In others a ratchet-like potential to simulate the many monomers translocation is used~\cite{shulten2004,linke2006}.

\begin{figure}[b]
 \centering
 \includegraphics[width=0.95\linewidth]{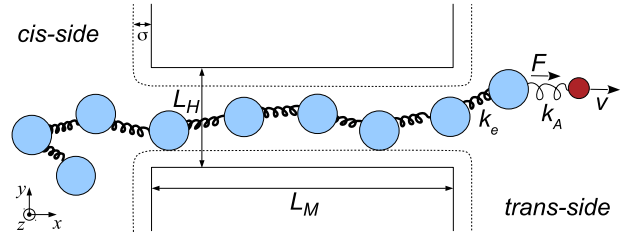}
 \caption{Section of the polymer translocating through a nanopore in the 3d space. The pore has a square section of width $L_H$ and its length is $L_M$, with the same repulsive walls as the whole membrane. The polymer is pulled through the pore with a force driven at constant velocity $v$.}
 \label{f-channel}
\end{figure}
Motivated by the results of different experiments with passive pores~\cite{KasPNAS96,Merchant2010,Schneider2010}, many studies of translocation have been performed, mainly under constant forces, either pore-driven or, less commonly, end-pulled~\cite{2017EPLSarabadani}. The role of active nanopores, with time dependent mechanisms assisting the translocation, has been only more recently considered. The stochastic opening or closing of the pore channel~\cite{2002Alberts,1998Petersen} is usually modelled by means of a dichotomous Markov noise (random telegraph noise)~\cite{ajf-rtn,ikonen2012,ajf-PRE2015}, and the pore activity is modelled as a sinusoidal pore actuation{~\cite{ajf-sin,golest2011,ikonen2012,golest2012}}. Sometimes, the translocation is assisted by ATP-fueled molecular motors~\cite{Bust01,Bust09}, modelled by specific stochastic forces~\cite{ajf-damn,pffs-damn,ajf-SR2017}.

Generally, the main magnitude studied in the translocation problem is the time spent by the polymer to cross the membrane and its scaling behavior with the polymer length. Another magnitude is the energy involved in the process, i.e. the free energy difference gained during the translocation in passing the barrier generated by the membrane walls. Instead, no much attention has been given to the force exerted to allow the translocation.
Experimentally, the study of the force has been recently attenctioned in many out of equilibrium processes by using the single molecule \emph{dynamic force spectroscopy}.  The method consists in the application of controlled forces to biological macromolecules or complexes \cite{Bhutan,Borgia2008,bust2004,Neumann2008}, by pulling one edge of the object at a constant velocity in order to induce conformational
changes -- such as the unfolding of proteins \cite{Puncher2009}, nucleic acids
\cite{Williams2002}, DNA secondary structures as the G-quadruplex \cite{baf-SR2017} -- or drive the dissociation of ligand-receptor complexes
\cite{Florin1994,Liphart2001,Rief1997,Merkel1999,falo2017}. The same velocity dependent end-pulling has been also implemented in some DNA translocation experiments \cite{2016Bulushev,2006NPKeyser,2010JPSischkal}. 

A number of models have been created in order to extract quantitative information from those velocity-dependent forces, all of them based in the thermal barrier escape. The Bell-Evans-Ritchie~\cite{Evans1997} approach is the first and more approximated one, where only the height of the barrier is modified by the applied force. More refined is the Dudko-Hummer-Szabo model~\cite{Dudko2006}, where various aspects of the potential profile are modelled, as barrier position, barrier height, and escape rate. Both the above models are build under the nontrivial hypothesis that no refolding events, or reattachments, occur in the phenomenon. So, they are useful to model systems pulled at relatively high velocities, when the refolding probability is relatively low. Differently, the Friddle-Noy-DeYoreo (FNY) approach~\cite{Yoreo2012} does take into account those refolding occurrences, making it suitable for modeling also slow driven events.

In this paper we study a set of translocation trajectories where the polymer is end-pulled at a constant velocity, and apply the FNY model to analyze the force recorded at different velocities, under the hypothesis that the whole translocation can be minimalistically described as the crossing of a free energy barrier, in analogue way as a rupture events. The possible  recombination events taken into account in this approach well describe the temporary -- eventually multiple -- return back to the \emph{cis} side of the monomer that entered the pore. With this method it is possible to evaluate at the same time the free energy barrier related to the translocation, and the limit force to apply in order to allow the very translocation.

\section{Models and system equations.}
The polymer is formed by $N$ identical monomers moving in the three-dimensional space, that  interact by means of elastic bonding, bending energy, excluded volume effects and repulsive interactions with both the membrane and the pore. The elastic potential energy is given by
 \be V_{\rm el}(d_i)=\frac{k_e}{2}\sum_{i=1}^{N} (d_i-d_0)^2,
 \label{v-har}
 \ee
\noindent where $k_e$ is the elastic parameter,
$d_i=|\bm{d}_i|=|\bm{r}_{i+1}-\bm{r}_i|$ is the distance between the monomers $i$ and $i+1$, with $\bm{r}_i$ the
position of the $i$-th particle and $d_0$ the equilibrium distance between consecutive monomers.
The bending energy of the chain is taken into account with the term
 \be
  V_{\rm ben}(\theta_i)=\frac{k_b}{2}\sum_{i=1}^{N} [1-\cos(\theta_i-\theta_0)],
 \label{v-ben}
 \ee
where $k_b$ is the bending elastic constant, $\theta_i$ is the
angle between the links $\bm{d}_{i+1}$ and $\bm{d}_{i}$, and $\theta_0$ the equilibrium angle, with $\theta_0=0$ in our case. With this term, our model is a discrete version of the worm-like chain (WLC) model.
In order to consider excluded volume effects between any couple of monomers, a repulsive only Lennard-Jones potential has been taken into account:
% \be V_{\rm
%   LJ}(r_{ij}) = 4 \epsilon \sum_{i \ne j=1}^{N} \left[ \left(\frac{\sigma}{r_{ij}}\right)^{12}-
%   \left(\frac{\sigma}{r_{ij}}\right)^6 \right]
% \label{LJ}
% \ee
 \be V_{\rm
   LJ}(r_{ij}) = 4 \epsilon \sum_{i \ne j=1}^{N} [( \frac{\sigma}{r_{ij}})^{12}-
   (\frac{\sigma}{r_{ij}})^6]
 \label{LJ}
 \ee
for $r_{ij}\leq 2^{1/6}\sigma$, and $-\epsilon$ otherwise, with $r_{ij}$ the distance between monomer $i$ and monomer $j$.
The dynamics of every monomer is obtained by the
overdamped Langevin equation of motion
 \bea
  m\gamma\dot{\bm{r}_i} = &-& \nabla_i V_{\rm el}(d_i)
-\nabla_i V_{\rm ben}(\theta_i) - \nabla_i V_{\rm
LJ}(r_{ij}) \nonumber \\
 &+& F_{drv} \bm{i} + \bm{F}_{sp,i} + \sqrt{2m\gamma k_BT} \, \bm{\xi}_i(t),
 \label{eq}
 \eea
where the effective viscosity parameter of each monomer is
included in the normalized time units. $\bm{\xi}_{i}(t)$ stands
for the Gaussian uncorrelated thermal fluctuation and follows the
usual statistical properties $\langle\xi_{i,\alpha}(t)\rangle=0$
and $\langle\xi_{i,\alpha}(t)\xi_{j,\beta}(t')\rangle = \delta_{i
j}\delta_{\alpha,\beta}\delta(t'-t)$, with $i=1,...,N$,
$\alpha, \beta = x,y,z$, and $\mathbf{\nabla}_i =
\partial / \partial x_i \, \bm{i} + \partial / \partial y_i
\, \bm{j}  + \partial / \partial z_i \, \bm{k}$.

The force term $\bm{F}_{sp}$ includes both the chain-membrane and chain-pore spatial constraint. This interaction force is modelled with the same repulsive Lennard-Jones potential described above. It takes place uniformly and perpendicularly to all the planes that define both the membrane and the pore channel, modelled as a square prism of base $L_M$ and length $L_H$ (see Fig.~\ref{f-channel}).

Finally, $F_{drv}$ is the driving force provided by an external source.
In accordance with experimental procedures, the cantilever move with constant velocity $v$, and pulls the potential through a quadratic potential, modelled as an interposed harmonic spring attached to the first monomer of the chain (See Fig.~\ref{f-channel}). The resulting force is
 $
   F_{drv}(x)  = -k_A (vt - l_A - x_N(t)),
   \label{PullF}
 $
then the loading force rate $r = dF(t)/dt \approx k_Av$. $k_A$ and $l_A$ are, respectively, the elastic constant and rest length of the spring, and $x_N(t)$ is the position of the first monomer of the chain, where the force is applied.
The use of the spring to apply the force to the polymer presents two advantages: From one hand it mimics the force applied to a molecule by an optical tweezer or an atomic force microscope commonly used in pulling experiments at constant velocity; On the other hand, it allows to record the value of the force applied in each instant of time, according to the polymer reaction to the pull and the pore interaction.

The analysis of the force trajectory is performed by using the FNY model, which describes the kinetic features of induced unfolding events (see Ref.\cite{Yoreo2012} for details). 
The basic idea of this model is that the unfolding process of a molecular complex can be depicted as an escape from a potential barrier, without additional complexities. The model predicts the presence of a limit pulling force that is reached when the probability to have the unfolded and the folded states are equal. In this case the folding rate $k_b(f)$ -- which decreases as the applied the force $f$ increases -- and the unfolding rate $k_u(f)$ -- which increases as $f$ increases -- have the same value. The unfolding rate $k_u(f)=k_0\exp(\beta (fx_u-0.5k_Ax_u^2))$ depends on the static escape rate $k_0$, the elastic constant of the harmonic cantilever $k_A$ and the position of the barrier $x_u$.

Specifically, the mean rupture/unfolding force predicted in this model is:
\begin{equation}
  \langle F_{u} \rangle= f_{eq}+ \frac{1}{\beta x_u}
  e^{\left(\frac{k_u(f_{eq})}{\beta rx_u}\right)} E_1 \left(\frac{k_u(f_{eq})}{\beta rx_u} \right),
  \label{eq:Yoreo}
\end{equation}
where $\gamma \approx 0.577$ is the Euler's constant, $\beta = 1/k_{\rm B}T$, and $E_1(z)=\int_z^{\infty} \frac{e^{-s}}{s} ds$ is the exponential integral which can be interpolated by the analytical function $e^zE_1(z)\approx \ln (1+e^{-\gamma}/z)$. 
The parameter $f_{eq}$ is the equilibrium force at which the folding and unfolding rates equal, and is related to the barrier hight $G^+$ as $f_{eq}=\sqrt{2k_AG^+}$.

The three parameters modeled in the translocation process are interpreted in an analogue way as the unfolding as follows: $x_u$ is the barrier position  in the reaction coordinate, $f_{eq}$ corresponds to the limit force to allow the translocation to occur, and $k_0$ represents the whole translocation rate.

\emph{Units and parameters.}
Following \cite{ikonen2012}, we define $m$, $d_0$, and $\epsilon_0$ as the mass, the length, and the energy units respectively. This choice determines a Lennard-Jones time scale given by $t_{LJ}= (md_0^2/\epsilon_0)^{1/2}$. However, as the
dynamics we propose is overdamped, the time scale that normalise the equation of motion Eq.~(\ref{eq}) is $t_{OD} = \gamma {t_{LJ}}^2 $, thus depending on the damping parameter. To set some values, let us
consider a DNA molecule at room temperature ($k_BT= 4.1\,\rm{pN nm}$)
and the simplest model with $k_b=0$. We have fixed our simulation
temperature to $k_BT=0.1$ in dimensionless units. This choice fixes our energy unit in $\epsilon_0= 41\,\rm{pN nm}$. By setting $d_0 = 1.875\,\rm{nm}$ and $m = 936\,\rm{amu}$ \cite{ikonen2012}, we obtain $t_{LJ} \approx 0.38$ps, while the force unit is given by $\epsilon_0/d_0 = 21.9\,\rm{pN}$. An
estimation for the kinetic damping is $\gamma \approx 1.6 \times
10^{13}\,\rm{s^{-1}}$, so obtaining $t_{OD} \approx 2.3\,\rm{ps}$. Other normalizations can be used depending on the system to simulate~\cite{af2012}.

We use a channel with a fixed length, $L_M=5.5 d_0$, longer than the distance of two consecutive monomer, and square section $L_H=2 d_0$. The rest distance between adjacent monomers is $d_0=1$ and $k_e = 1600$, large enough to maintain the
bonds of the chain rigid. The Lennard-Jones energy is $\epsilon=0.3$, and $\sigma=0.88$. The values of $d_0$, $\sigma$, $L_M$ and $L_H$ guarantees that the polymer is maintained almost linear and ordered inside the pore. Also, the different choices of the bending constant $k_b$, gives the possibility to study the magnitudes for different \emph{persistence lengths} of the chain, i.e. the stiffness of the polymer. For our model $L_p=k_b/k_BT$. Thus for example we obtain $L_p=5 d_0$ for $k_b=0.5$. 

\begin{figure}[t]
\includegraphics[angle=-90, width=0.95\linewidth,left]{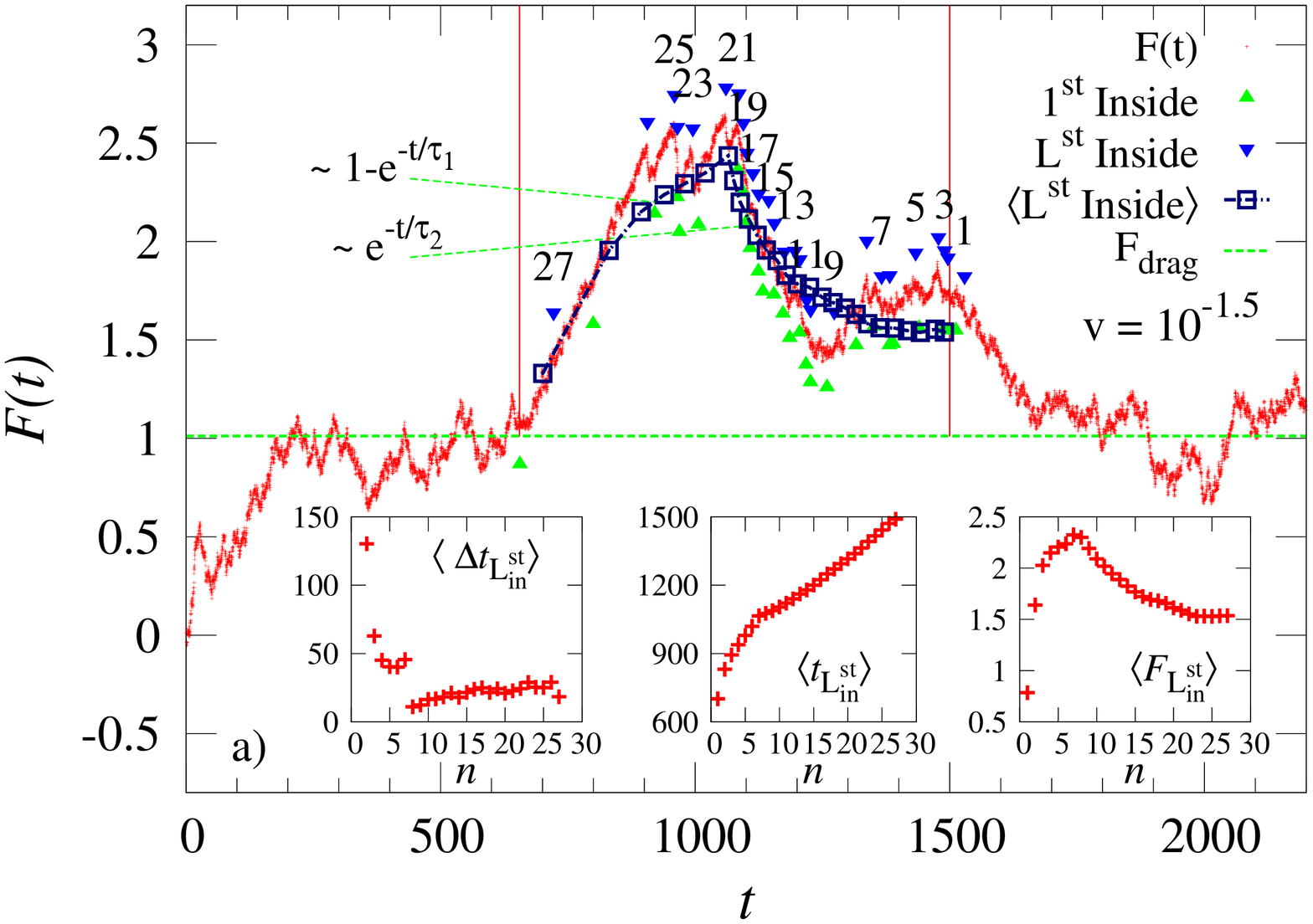}
\hspace{1cm}\includegraphics[angle=-90, width=0.97\linewidth,right]{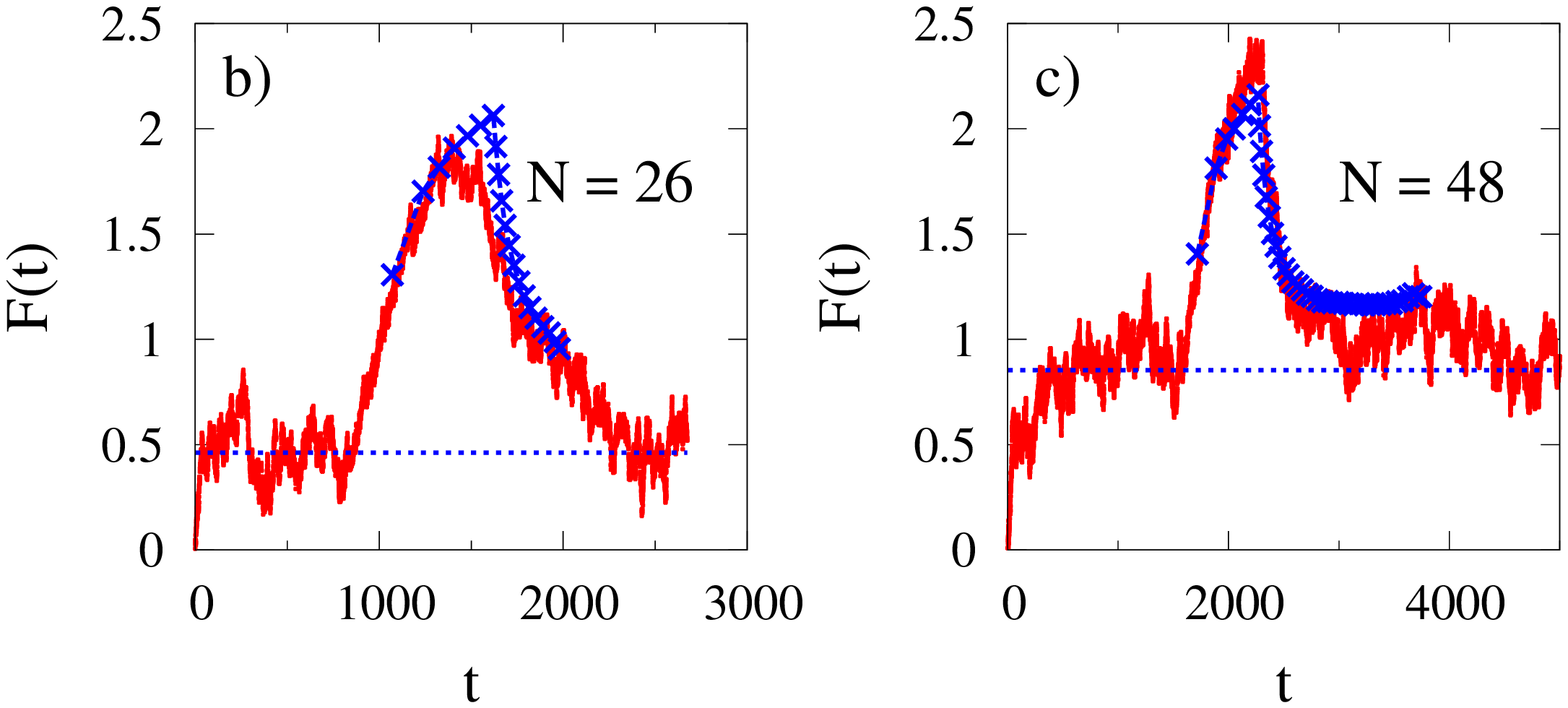}
\caption{a) Force trajectory as a function of time for $N=32$ monomers, $v=10^{-1.5}$, bending $k_b = 10^{-3}$. The small arrows above and below the curve indicate, respectively, the time of the last and first entrance inside the pore of the monomer (index label above the upper points). The square symbols indicate the average value of the force at the entrance events (last entrance in case of multiple entrance-exits). The horizontal line represents the value of the drag force $F_{drag}$ in the absence of both pore interaction and fluctuations. The two vertical lines define the time interval where the mean force has been calculated. The three insets report, respectively, the mean waiting entrance times, the total time spent to enter, the mean force at the last entrance, as a function of the number of monomers that enter the pore $n$. b) c): two examples of $F(t)$ with $N=24$ and $N=48$.}
 \label{f-Ft}
 \end{figure}

\emph{Definitions and simulation details.}
We span over different velocities $v$ of the cantilever. Each of the $N_{exp}=300$ simulations start with all the monomers lying linearly ordered along the $x$-axis at the rest equilibrium distance, with the monomer closest to the pore far away from the pore entrance.
During a thermalization time $t_t = 1000$t.u. the chain evolves under the action of thermal fluctuations in order to reach a thermalized state according to the temperature. After that transient time, the full dynamics given by Eqs.~(\ref{eq}) acts, and, because of the  application of the pull force, the polymer start moving directed to the pore, while keeping fixed the position of the firsts five monomers, in order to drive their entrance inside the pore without initial wall reactions. The force of the spring is monitored along the whole dynamics.  
When the $i+1$ trajectory starts, the initial polymer configuration is the one already thermalized in the $i$-th realization. Then it undergoes anew the thermalization which again lasts a time $t_t$. This way the polymer continues thermalizing before starting each new trajectory.

This setup permits to consider the trajectories where the polymer enters the pore having already reached the constant drag velocity equal to the pull velocity $v$. This way the exceeded force registered during the translocation (region between the two vertical lines in Fig.~\ref{f-Ft}) is only due to the interaction pore-polymer, and its values can be related to the pore free energy barrier.

\section{Results. }
A typical $F(t)$ trajectory is reported in Fig.~\ref{¼f-Ft}a), whose inspection evidences a rich phenomenology. The small arrows above the force trajectories indicate the last pore-entrance events done by the monomers indicated with their indexes (the labels above the arrows). The events reported are the last ones not followed by exit back-steps recorded between the first entrance (small arrows below the curve) and the last. Interestingly the average over the number of realizations of the last entrance events in both times and forces (joined squares in the center of the figure) reveals the existence of two distinct regimes. The force increase exponentially up to the entrance of the seventh monomer, and then decreases, again exponentially, up to the exit of all the chain. This behavior repeats for all the velocities considered, and at different chain lengths. This change of regime is also visible in the three insets of the figure, where various magnitudes are shown as a function of the number $n$ of the monomers that entered the pore. Specifically, from left to right, the mean waiting entrance times $\langle \Delta t_{L_{in}^{st}} \rangle$, \emph{i.e.} the time before a monomer enters for the last time in the pore counted from the time of the last entrance of the previous one, the total mean entrance times $\langle t_{L_{in}^{st}} \rangle$, \emph{i.e.} the total time before a monomer enters the pore, and the mean entrance forces $\langle F_{L_{in}^{st}} \rangle$, \emph{i.e.} the mean force registered at the spring when each monomer enters the pore. This change of regime is due to the fact that, once reached the \emph{trans} region, the entropic contribution due by the interaction with the walls by the \emph{trans} monomers start helping the translocation, as the interaction with the walls due to the \emph{trans} monomers results in a force pushing in the same direction of the pulling. This feature also explains the value of the number of monomers involved in the force decrease ($7$ monomers), which only depends of the length of the pore $L_M$, in the different conditions simulated. In fact, different chain lengths [we tried $N=26$ and $N=48$, see Figs.~\ref{f-Ft}(a), and \ref{f-Ft}(b)] present the same number of monomers before the maximum value of the force. It is worth to remember that the firsts five monomers are held fixed in the transversal directions also while they move in the \emph{trans} region, in order to maintain the same conditions before, after, and during the pore crossing.

\begin{figure}[b]
\centering
\includegraphics[angle=-90, width=0.95\linewidth]{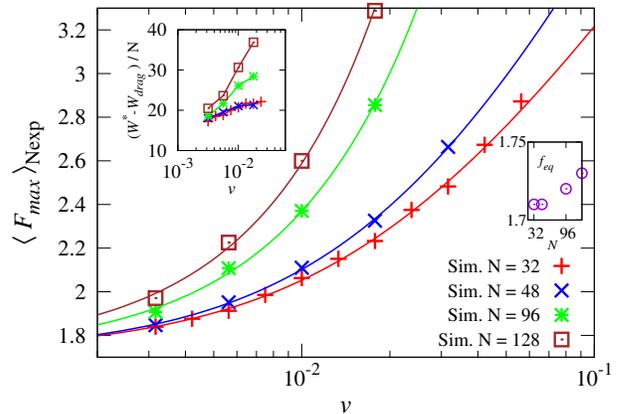}
\caption{Average value of the maximum force $F_{max}$ registered at the spring \emph{during} the translocation as a function of the pulling velocity $v$ for $N=32$, $N=48$, $N=96$, and $N=128$  with $k_A=0.2$, $k_b=10^{-3}$ (symbols), with the corresponding FNY fit (full lines). Left inset: The work $W^*$ transferred by the cantilever without the work to drag the polymer ($W_{drag}$) per translocated monomer. Right inset: the limit equilibrium force $f_{eq}$ as a function of $N$.}
 \label{f-Fv}
 \end{figure}

\begin{table}
\begin{center}
\begin{tabular}{c c c c  c | c}
\hline \hline
                 &           &    N     &          &           &   \\
                 &   32    &    48    &  96    &  128   &  Error \\
\hline
$x_u$    &  0.11  &  0.08  &  0.0047 &  -0.008           &   $\approx 80\% $\\
$k_0$    &  0.0027  &  0.0024  &  0.0017 &  0.0013  &  $\approx  7\%   $\\
$f_{eq}$ &  1.71   &  1.71   &  1.72    &  1.73             &   $\approx 2\%   $\\
$G^+$    &  7.34   &  7.34   &  7.35   &  7.36     \\
%\hline
\end{tabular}
\end{center}
\caption{Values of the FNY fit parameters $x_u$, $k_0$, $f_{eq}$, and the related estimated barrier $G^+$ for different values of the polymer length ($N$). The last column reports the order of magnitude of the error in the fit.}
\label{table}
\end{table}

\emph{Reaction of the walls.}
The energy barrier in the translocation is essentially due to the interactions of the polymer with the membrane walls that push in the opposite direction of the movement. This contribution is not easy to evaluate directly, and we use here a force spectroscopy approach which consists in measuring the mean force recorded during the translocation at the pulling spring.

Fig.~\ref{f-Fv} shows the ensemble average over $N_{exp}$ of the maximum force $\langle F_{max} \rangle$ registered at the spring from the \emph{first entrance} of the first monomer up to the \emph{last entrance} of the last one (See Fig~\ref{f-Ft}), as a function of the pulling velocity for different polymer lengths $N=32$, $N=48$, $N=96$, and $N=128$, with the bending parameter $k_b=10^{-3}$.
The full lines in the figure reports the fits to the calculated points according to the FNY model, showing an excellent agreement with the data. The model (see Eq.~\ref{eq:Yoreo}) provides a quantitative estimation of the most relevant magnitudes involved in the escape from a potential well whose values have been resumed in Table~\ref{table}.
The fitted parameter $x_u$ represents the position of the barrier in the reaction coordinate of the system. It has values close to zero in all the cases. This position could be located in our model at the pore entrance, that lies at the coordinate $x=0$.

The inspection of Fig.~\ref{f-Fv} shows a similar behavior for the different polymer lengths. One of the results in this data is the presence of a clear limit force $f_{eq}$ (See Table~\ref{table}), as demonstrate the saturating trend of the curves for low velocities. This force is the minimum force necessary to apply for the translocation to occur, i.e. to overpass the stochastic wall reaction due to the impacts of the polymer in the \emph{cis} side of the membrane which oppose the rightward movement. Even if this force could be expected higher for longer polymers, the actual values registered in this approach is weakly dependent on the length we have simulated. The values are reported in the resume plot which shows $f_{eq}$ $vs$ the number of monomers $N$, plotted in the right inset of Fig.~\ref{f-Fv}. This weak dependence with the polymer length $N$ has been also confirmed by the stall force evaluation presented below in this paper.
The presence of a threshold force that allows the translocation has been also observed in previous calculations \cite{ajf-PRE2015,MenaisPRE2018}. 
 \begin{figure}[t]
   \begin{tabular}{c}
    \hspace{0.5cm}\includegraphics[width=.43\linewidth,left]{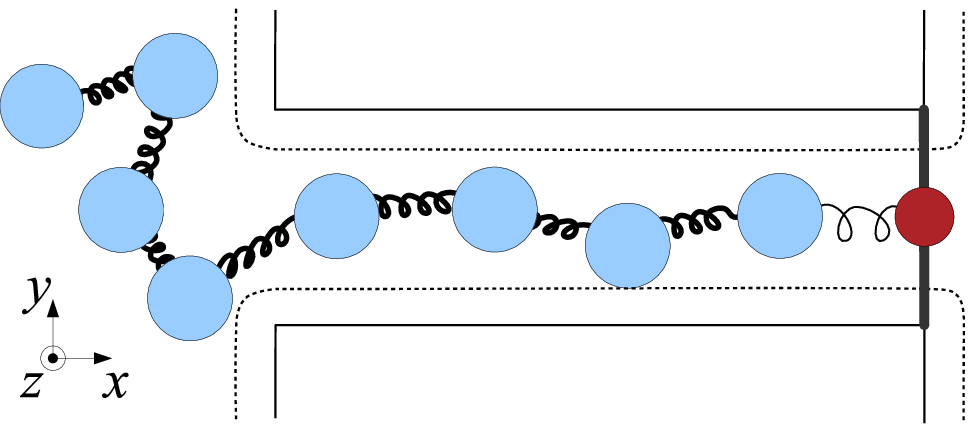}
        \\[-2.4cm]
    \includegraphics[angle=-90, width=.99\linewidth]{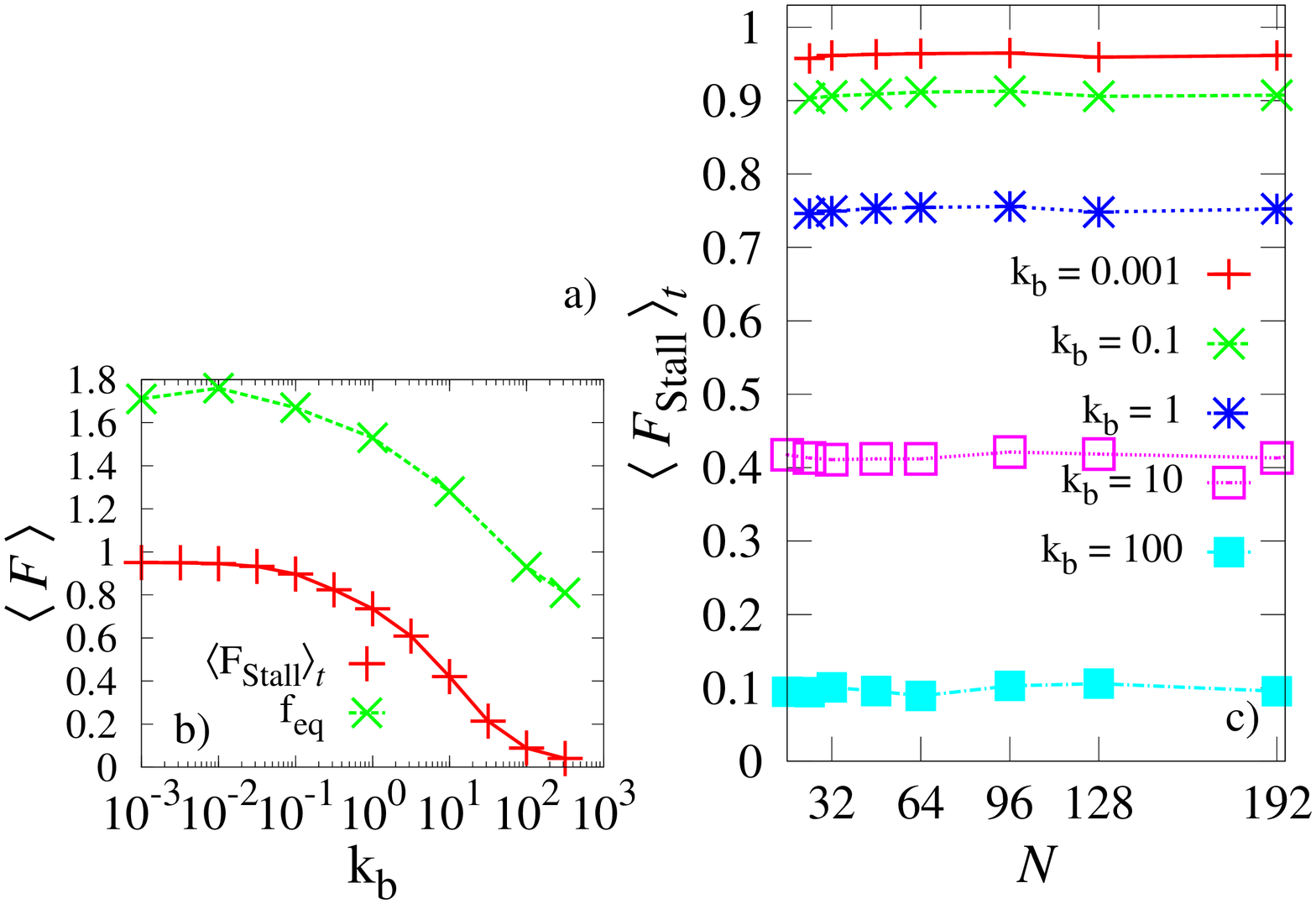}
   \end{tabular}
   \label{f-stack-2}
   \caption{a) Stall force setup. The cantilever edge remains fixed at the pore exit while the dynamics runs. b) Mean value of the stall force registered at the spring $\langle F_{stall}\rangle_t$ and the limit force $f_{eq}$ calculated through the FNY fit as a function of the bending constant $k_b$, with $N=32$. c) $\langle F_{stall}\rangle_t$ as a function of the chain length for different bending parameters from $k_b=10^{-3}$ (i.e. with persistence length $L_p \approx 0$) to $k_b=10^{2}$ ($L_p \approx 1000$).}
   \label{f-stack}
\end{figure}

The left inset of Fig.~\ref{f-Fv} reports the calculation of the work per monomer transferred to the polymer by the cantilever. This work has been calculated for each trajectory as $W_m=\int_{x_N(t_{in})}^{x_N(t_{out})} F dx/N - W_{drag}$, averaged over the number of realizations $N_{exp}$ by using the Jarzynski equality~\cite{Jarzy1997}, i.e. $W^* =-1/\beta \ln \langle e^{- \beta W_m} \rangle$. The subtracted term $W_{drag}=F_{drag} [x_N(t_{in})-x_N(t_{out})]/N$ is the estimated work to drag the polymer, subtracted from the total work done by the cantilever $\int_{x_N(t_{in})}^{x_N(t_{out})} F dx/N$. The time thresholds $t_{in}$ and $t_{out}$ represent the times of the entrance of the first monomer and the exit of the last one, respectively. The value $W^*$ so calculated is then the Helmholtz free energy difference $\Delta F$ between the \emph{trans} and the \emph{cis} side, difference that takes into account that the two states are not at the same level given the directionality imposed by the pulling.

\subsubsection{Stall forces.}
In principle, the equilibrium value $f_{eq}$ obtained with the FNY fit can be compared with the force calculated with the following different method: A set of calculations has been performed by maintaining the spring at the \emph{trans} edge of the pore channel, and measuring the average force registered at the spring, which is so the necessary one to hold the chain at a fixed position (scheme in panel a) of Fig.~\ref{f-stack}).
Panel b) of Fig.~\ref{f-stack} shows the monotonic decrease of the mean stall force $\langle F_{stall}\rangle_t$ as a function of the bending parameter $k_b$. This means that, in static conditions, a lower force is required to maintain inside the pore a chain with higher bending values. This is due to the fact that -- given the directionality imposed by the extended pore -- a stiffer polymer presents a lower interaction with the membrane walls, so making a smaller force able to hold the chain in those cases.
The values of $f_{eq}$ evaluated by means of the FNY fit have also been plotted for different bending values in the same panel b). The two behaviors have a very similar trend but the limit force $f_{eq}$ is higher that the stall force. While the $\langle F_{stall}\rangle_t$ is measured with fixed conditions of the cantilever, the dynamical origin of the $f_{eq}$ (measured during the translocation) make those two measures slightly different each other, tough correlated. In the first case the mean force $\langle F_{stall}\rangle_t$ is the necessary one to \emph{maintain} the chain attached to the pore, in the other case $f_{eq}$ is the \emph{limit force} evaluated when the translocation actually occurs. So, the two forces are expected to be similar, though not identical, even if the differences are almost constant at all bending values.

Fig.~\ref{f-stack} (right panel c)) reports the results of the mean stall force $\langle F_{stall}\rangle_t$ as a function of the number of monomers $N$, for different values of the bending parameter $k_b$. For each of the value of the bending parameter, the stall force curve, after a small increase with the polymer length, tends to maintain independent on the number of monomers. This results, apparently counter-intuitive, can be explained with the picture that, even if the higher number of monomers should give a higher interaction with the membrane walls, and so a higher stall force, the longer the polymer is, the higher is the distance of the polymer centre form the membrane, so decreasing the number of interactions with the wall due to the peripheral monomers. The weak dependence of the stall force with the number of monomers $N$ appears to confirm the behavior reported in the right inset of Fig.~\ref{f-Fv}, where the limit force $f_{eq}$ obtained from the FNY fit shows a low dependence with the polymer length.
As concerns the rigidity, as the bending parameter increases, the stall force decreases monotonically, as it may be expected. The main qualitative ingredients of these results -- \emph{i.e} the decrease of the force with the persistence length, and the constant trend of the force with the polymer length-- have been also obtained in Ref.~\cite{2003Vicsek,2011Rowghanian} where the authors find an analytical expression able to reproduce those behaviors. 
It is worth to note that a power law scaling with the polymer size has been reported in \cite{Mondaini2014}, where the entropic barrier has been studied with a chain model similar to that used in this paper, with the use of extensible bonds.

\section{Summary and Conclusions.}
We have studied the translocation of a polymer chain pulled by a cantilever moving at a constant velocity through a uniformly repulsive pore membrane in a fluctuating environment.
The dependence of the average of the maximum translocation force as a function of the velocity has been calculated and its analysis with the FNY approach has provided a quantitative characterization of the limit force allowing the translocation, as well as the potential barrier according to the FNY model. The stall force to hold one edge of the chain in the pore has been calculated in static conditions with a different approach, and we found a similar decreasing behavior with the rigidity of the chain as the FNY outcomes. This work constitutes an explorative approach to the force spectroscopy analysis in polymer translocation, which appears to be a promising method to face the richness of the translocation problem.

\section*{Acknowledgments}
This work is supported by the Spanish projects MINECO No.~FIS2017-87519-P and No.~FIS2014-55867-P, both co-financed by the Fondo Europeo de Desarrollo Regional (FEDER) funds. We also acknowledge the support of the Arag\'on Government grant No. E19 to the FENOL group.

\end{document}